\documentclass[prc,twocolumn,showpacs,amsmath,amssymb,superscriptaddress,nofootinbib]{revtex4-1}

\usepackage[utf8]{inputenc}
\usepackage{epsfig}
\usepackage{graphicx}
\usepackage{dcolumn}
\usepackage{bm}
\usepackage{color}
\usepackage{url}
\usepackage{multirow}
\usepackage{enumitem}

\newcommand{ \pT} {p_{\rm T}}

\newcommand{\sNN}{$\sqrt{s_{\rm NN}}$ }

\newcommand{\GeVc}{GeV/$c$}
\newcommand{\GeVcsq}{GeV/$c^2$}

\newcommand {\gevc}	{GeV/$c$}

\newcommand {\pdA}	{$p(d)$+Au}

\newcommand {\pPb}	{$p$+Pb}
\newcommand {\PbPb}	{Pb+Pb}

\newcommand {\phia}	{\phi_{\alpha}}
\newcommand {\phib}	{\phi_{\beta}}
\newcommand {\phic}	{\phi_c}

\newcommand {\vc}	{v_{2,c}}

\newcommand {\vres}	{v_{2,{\rm res}}}
\newcommand {\gOS}	{\gamma_{\rm OS}}
\newcommand {\gSS}	{\gamma_{\rm SS}}
\newcommand {\dg}	{\Delta\gamma}

\newcommand {\phires}	{\phi_{\rm res}}

\newcommand {\minv} {m_{\rm inv}}

\newcommand {\mean}[1]	{\langle #1\rangle}

\begin{document}
\title{Pair invariant mass to isolate background in the search for the chiral magnetic effect in Au+Au collisions at $\sqrt{s_{_{\rm NN}}}$= 200 GeV}

\affiliation{Abilene Christian University, Abilene, Texas   79699}
\affiliation{AGH University of Science and Technology, FPACS, Cracow 30-059, Poland}
\affiliation{Alikhanov Institute for Theoretical and Experimental Physics NRC "Kurchatov Institute", Moscow 117218}
\affiliation{Argonne National Laboratory, Argonne, Illinois 60439}
\affiliation{American University of Cairo, New Cairo 11835, New Cairo, Egypt}
\affiliation{Brookhaven National Laboratory, Upton, New York 11973}
\affiliation{University of California, Berkeley, California 94720}
\affiliation{University of California, Davis, California 95616}
\affiliation{University of California, Los Angeles, California 90095}
\affiliation{University of California, Riverside, California 92521}
\affiliation{Central China Normal University, Wuhan, Hubei 430079 }
\affiliation{University of Illinois at Chicago, Chicago, Illinois 60607}
\affiliation{Creighton University, Omaha, Nebraska 68178}
\affiliation{Czech Technical University in Prague, FNSPE, Prague 115 19, Czech Republic}
\affiliation{Technische Universit\"at Darmstadt, Darmstadt 64289, Germany}
\affiliation{ELTE E\"otv\"os Lor\'and University, Budapest, Hungary H-1117}
\affiliation{Frankfurt Institute for Advanced Studies FIAS, Frankfurt 60438, Germany}
\affiliation{Fudan University, Shanghai, 200433 }
\affiliation{University of Heidelberg, Heidelberg 69120, Germany }
\affiliation{University of Houston, Houston, Texas 77204}
\affiliation{Huzhou University, Huzhou, Zhejiang  313000}
\affiliation{Indian Institute of Science Education and Research (IISER), Berhampur 760010 , India}
\affiliation{Indian Institute of Science Education and Research (IISER) Tirupati, Tirupati 517507, India}
\affiliation{Indian Institute Technology, Patna, Bihar 801106, India}
\affiliation{Indiana University, Bloomington, Indiana 47408}
\affiliation{Institute of Modern Physics, Chinese Academy of Sciences, Lanzhou, Gansu 730000 }
\affiliation{University of Jammu, Jammu 180001, India}
\affiliation{Joint Institute for Nuclear Research, Dubna 141 980}
\affiliation{Kent State University, Kent, Ohio 44242}
\affiliation{University of Kentucky, Lexington, Kentucky 40506-0055}
\affiliation{Lawrence Berkeley National Laboratory, Berkeley, California 94720}
\affiliation{Lehigh University, Bethlehem, Pennsylvania 18015}
\affiliation{Max-Planck-Institut f\"ur Physik, Munich 80805, Germany}
\affiliation{Michigan State University, East Lansing, Michigan 48824}
\affiliation{National Research Nuclear University MEPhI, Moscow 115409}
\affiliation{National Institute of Science Education and Research, HBNI, Jatni 752050, India}
\affiliation{National Cheng Kung University, Tainan 70101 }
\affiliation{Nuclear Physics Institute of the CAS, Rez 250 68, Czech Republic}
\affiliation{Ohio State University, Columbus, Ohio 43210}
\affiliation{Institute of Nuclear Physics PAN, Cracow 31-342, Poland}
\affiliation{Panjab University, Chandigarh 160014, India}
\affiliation{Pennsylvania State University, University Park, Pennsylvania 16802}
\affiliation{NRC "Kurchatov Institute", Institute of High Energy Physics, Protvino 142281}
\affiliation{Purdue University, West Lafayette, Indiana 47907}
\affiliation{Rice University, Houston, Texas 77251}
\affiliation{Rutgers University, Piscataway, New Jersey 08854}
\affiliation{Universidade de S\~ao Paulo, S\~ao Paulo, Brazil 05314-970}
\affiliation{University of Science and Technology of China, Hefei, Anhui 230026}
\affiliation{Shandong University, Qingdao, Shandong 266237}
\affiliation{Shanghai Institute of Applied Physics, Chinese Academy of Sciences, Shanghai 201800}
\affiliation{Southern Connecticut State University, New Haven, Connecticut 06515}
\affiliation{State University of New York, Stony Brook, New York 11794}
\affiliation{Instituto de Alta Investigaci\'on, Universidad de Tarapac\'a, Arica 1000000, Chile}
\affiliation{Temple University, Philadelphia, Pennsylvania 19122}
\affiliation{Texas A\&M University, College Station, Texas 77843}
\affiliation{University of Texas, Austin, Texas 78712}
\affiliation{Tsinghua University, Beijing 100084}
\affiliation{University of Tsukuba, Tsukuba, Ibaraki 305-8571, Japan}
\affiliation{Valparaiso University, Valparaiso, Indiana 46383}
\affiliation{Variable Energy Cyclotron Centre, Kolkata 700064, India}
\affiliation{Warsaw University of Technology, Warsaw 00-661, Poland}
\affiliation{Wayne State University, Detroit, Michigan 48201}
\affiliation{Yale University, New Haven, Connecticut 06520}

\author{M.~S.~Abdallah}\affiliation{American University of Cairo, New Cairo 11835, New Cairo, Egypt}
\author{J.~Adam}\affiliation{Brookhaven National Laboratory, Upton, New York 11973}
\author{L.~Adamczyk}\affiliation{AGH University of Science and Technology, FPACS, Cracow 30-059, Poland}
\author{J.~R.~Adams}\affiliation{Ohio State University, Columbus, Ohio 43210}
\author{J.~K.~Adkins}\affiliation{University of Kentucky, Lexington, Kentucky 40506-0055}
\author{G.~Agakishiev}\affiliation{Joint Institute for Nuclear Research, Dubna 141 980}
\author{I.~Aggarwal}\affiliation{Panjab University, Chandigarh 160014, India}
\author{M.~M.~Aggarwal}\affiliation{Panjab University, Chandigarh 160014, India}
\author{Z.~Ahammed}\affiliation{Variable Energy Cyclotron Centre, Kolkata 700064, India}
\author{I.~Alekseev}\affiliation{Alikhanov Institute for Theoretical and Experimental Physics NRC "Kurchatov Institute", Moscow 117218}\affiliation{National Research Nuclear University MEPhI, Moscow 115409}
% ...

\collaboration{STAR Collaboration}\noaffiliation

\date{\today}

\begin{abstract}
	Quark interactions with topological gluon configurations can induce local chirality imbalance and parity violation in quantum chromodynamics, which can lead to the chiral magnetic effect (CME) -- an electric charge separation along the strong magnetic field in relativistic heavy-ion collisions. The CME-sensitive azimuthal correlator observable ($\Delta\gamma$) is contaminated by background arising, in part, from resonance decays coupled with elliptic anisotropy ($v_{2}$). We report here differential measurements of the correlator as a function of the pair invariant mass ($m_{\rm inv}$) in 20-50\% centrality Au+Au collisions at $\sqrt{s_{_{\rm NN}}}$= 200 GeV by the STAR experiment at RHIC. Strong resonance background contributions to $\Delta\gamma$ are observed. At large $m_{\rm inv}$ where this background is significantly reduced, the $\Delta\gamma$ value is found to be significantly smaller. An event-shape-engineering technique is deployed to determine the $v_{2}$ background shape as a function of $m_{\rm inv}$. We extract a $v_2$-independent and $m_{\rm inv}$-averaged signal $\Delta\gamma_{\rm sig}$ 
= (0.03 $\pm$ 0.06 $\pm$ 0.08) $\times10^{-4}$, or $(2\pm4\pm5)\%$ of the inclusive $\Delta\gamma(m_{\rm inv}>0.4$ GeV/$c^2$)$ =(1.58 \pm 0.02 \pm 0.02) \times10^{-4}$, within pion $p_{T}$ = 0.2 - 0.8~\gevc and averaged over pseudorapidity ranges of
$-1 < \eta < -0.05$ and $0.05 < \eta < 1$. 
This represents an upper limit of $0.23\times10^{-4}$, or $15\%$ of the inclusive result, at $95\%$ confidence level
for the $m_{\rm inv}$-integrated CME contribution. 

\end{abstract}
\maketitle


\section{Introduction}
Quark interactions with topological gluon fields can induce chirality imbalance and local parity violation in quantum chromodynamics (QCD)~\cite{Lee:1974ma,Kharzeev:1998kz,Kharzeev:1999cz}. 
This can lead to electric charge separation in the presence of a strong magnetic field, 
a phenomenon known as the chiral magnetic effect (CME)~\cite{Fukushima:2008xe,Muller:2010jd}. 
Such a strong magnetic field is likely present in non-central heavy-ion collisions, 
mainly generated by the spectator protons~\cite{Kharzeev:2007jp,Asakawa:2010bu}, 
and may last an extended period of time~\cite{Kharzeev:2009pj,Tuchin:2013apa}. 
It has been suggested that the CME correlation signal can be observable in heavy-ion collisions~\cite{Kharzeev:2004ey,Shi:2017cpu} at the Relativistic Heavy-Ion Collider (RHIC)
and the Large Hadron Collider (LHC)~\cite{Kharzeev:1999cz,Kharzeev:2007jp}. 
Extensive efforts have been devoted to the search for a CME-induced charge separation along the magnetic field in heavy-ion collisions (see reviews in Refs.~\cite{Kharzeev:2015znc,Zhao:2018ixy,Zhao:2018skm,Zhao:2019hta,Li:2020dwr}). 
Many analysis techniques are being pursued~\cite{Sirunyan:2017quh,Acharya:2017fau,Xu:2017qfs,Tang:2019pbl}, 
and a CME-motivated isobar collision program was conducted at RHIC in 2018~\cite{Skokov:2016yrj,STAR:2021mii}.

In non-central, i.e.~finite impact parameter ($b$), heavy-ion collisions, 
the magnetic field is, on average, perpendicular to
the reaction plane (defined by the impact parameter direction and the beam). 
A surrogate for the reaction plane is the participant plane~\cite{Alver:2006wh}, which, in turn, can  
be estimated using the second-order harmonic plane ($\psi_2$) from the azimuthal distribution of final-state particles. 
Because topological fluctuations are random, the single particle asymmetry resulting from the charge separation vanishes. 
One needs to resort to two-particle correlations, a common observable of which is the three-point correlator with respect to $\psi_2$~\cite{Voloshin:2004vk}:
\begin{equation}
    \gamma\equiv\mean{\cos(\phia+\phib-2\psi_2)}\,,
\end{equation} 
where $\phia$ and $\phib$ are the azimuthal angles of particles $\alpha$ and $\beta$, respectively, 
either of same-sign (SS) or opposite-sign (OS) electric charges.
The CME would result in SS pairs close in azimuth and OS pairs 
back-to-back, both perpendicular to $\psi_2$, yielding $\gamma_{\rm SS}=-1$ and $\gamma_{\rm OS}=+1$.

\begin{figure}[hbt]
	\begin{center}
		\includegraphics[width=0.4\textwidth]{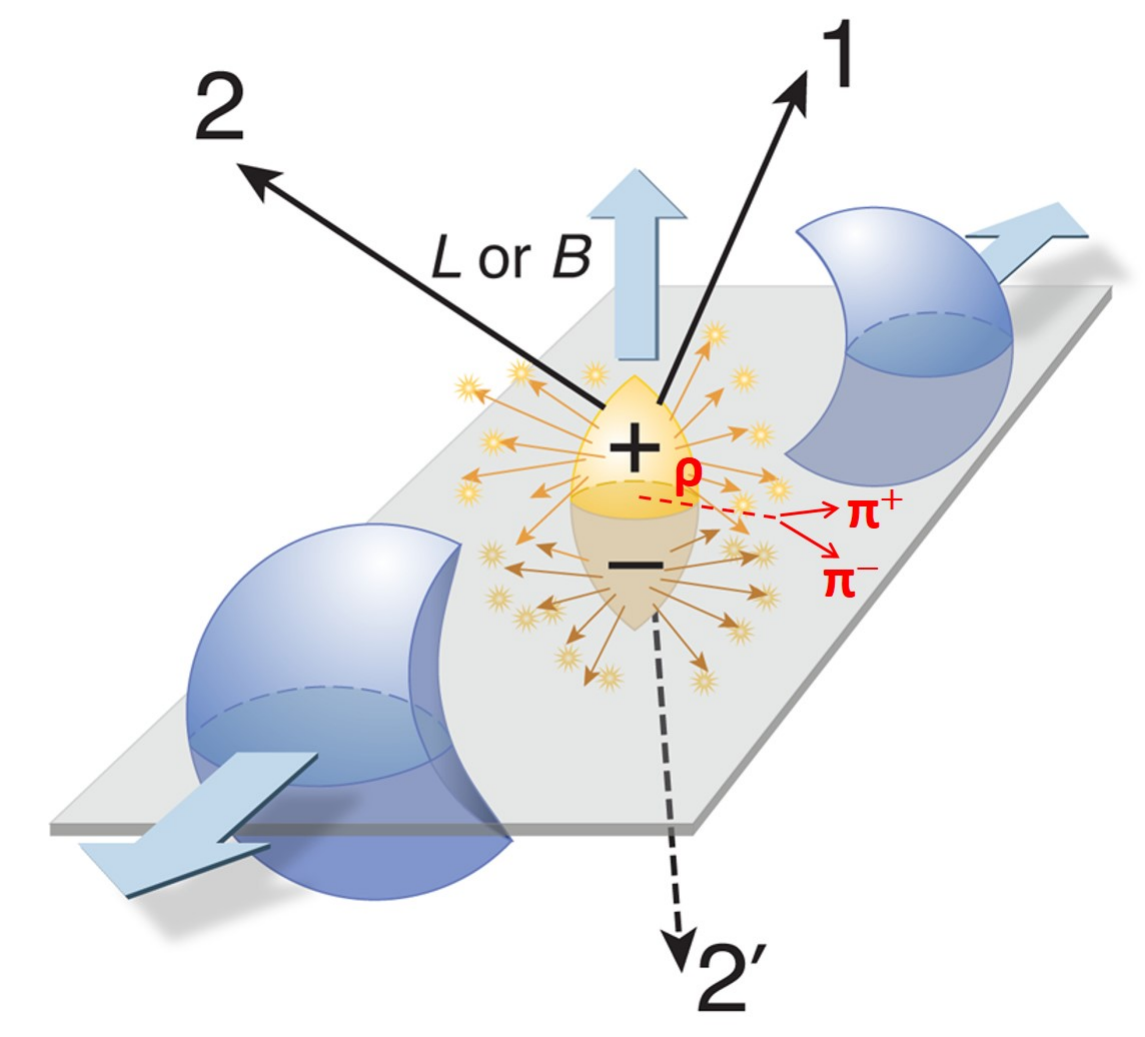}
		\caption{
			Schematic view of the charge separation along the system orbital momentum ($\vec{L}$),
			which coincides with the magnetic field ($\vec{B}$) direction. 
            The arrows 1 and 2 represent a SS pair close in azimuth, and 1 and 2' represent a OS pair ``back-to-back''.
            The $\rho$ resonance at $\psi_2$
            decays into a $\pi^+\pi^-$ pair, giving a positive $\dg_{\rm OS}$.
			This illustration is based on a figure in Ref.~\cite{Physics.2.104}.
			}
		\label{fig:cartoon}
	\end{center}
\end{figure}

It has been argued, prompted by data measurements~\cite{Abelev:2009ad,Abelev:2009ac}, that it is possible that the OS pair correlations are lost because of medium interactions and some of the SS pairs can still survive~\cite{Kharzeev:2007jp}. 
On the other hand, the underlying event could have charge-independent and charge-dependent correlations from non-CME physics.
These backgrounds can also alter the OS and SS correlations in such a way that they are not symmetric about zero any more.
To remove the charge-independent background (e.g. from global momentum conservation), 
the correlator difference, 
\begin{equation}
    \dg\equiv\gOS-\gSS\,,
    \label{eq:dg}
\end{equation}
is used~\cite{Voloshin:2004vk}.
A CME signal would yield a measurement of $\Delta\gamma>0$, 
the magnitude of which would be diluted by non-CME pairs. However, charge-dependent background correlations also exist, 
such as those from resonance decays~\cite{Voloshin:2004vk,Wang:2016iov}.
This is illustrated schematically in Fig.~\ref{fig:cartoon} using $\rho\rightarrow\pi^+\pi^-$ as an example, where the gray plane indicates the reaction plane and the total orbital angular momentum and the magnetic field are, on average, perpendicular to the reaction plane. 
Because more particles/resonances are produced parallel to $\psi_2$ than perpendicular, 
as quantified by the elliptic flow anisotropy parameter $\vres$,
the overall effect is a positive background $\dg_{\rm bkgd}>0$.
This flow-induced background can be expressed as: 
$\dg_{\rm bkgd} \propto \mean{\cos(\phia+\phib-2\phires)}\vres$,
where $\phires$ is the resonance azimuth, and $\alpha$ and $\beta$ are the resonance decay daughters~\cite{Voloshin:2004vk,Zhao:2019hta,Wang:2016iov,Schlichting:2010qia,Bzdak:2010fd}. 

Positive $\dg$ is indeed observed at the level of $\sim 10^{-4}$
in mid-central heavy-ion collisions where the two nuclei partially overlap at intermediate $b$ values~\cite{Abelev:2009ad,Abelev:2009ac,Adamczyk:2014mzf,Abelev:2012pa,Khachatryan:2016got}. 
A difficulty in its CME interpretation 
is the large charge-dependent background 
aforementioned, the magnitude of which dominates or may even fully account for the measured $\dg$
~\cite{Wang:2009kd,Bzdak:2009fc,Schlichting:2010qia,Asakawa:2010bu,Pratt:2010zn,Adamczyk:2013kcb,Khachatryan:2016got,Sirunyan:2017quh,Zhao:2019hta,Zhao:2019kyk}. 
The first experimental demonstration of such a background comes from
proton-lead collisions, where the participant plane is determined purely by geometry fluctuations (e.g., the proton strikes several nucleons in the lead nucleus, giving an irregular overlap shape), 
essentially uncorrelated with the impact parameter or the magnetic field direction~\cite{Khachatryan:2016got}. 
Any CME signal is expected to be negligible in small systems,
and yet, a large $\dg$ was observed in \pPb\ collisions at the LHC, similar to that in \PbPb\ collisions. 
This challenged the CME interpretation of the heavy-ion data~\cite{Khachatryan:2016got}.
A large $\dg$ is also observed in \pdA\ collisions at RHIC~\cite{STAR:2019xzd}.
Event shape engineering (ESE)~\cite{Schukraft:2012ah}, where events are selected within the same centrality bin but differing in $v_2$, has been used at the LHC to derive upper limits on the CME~\cite{Sirunyan:2017quh,Acharya:2017fau}. 
The CME signal likely depends on the collision energy, but quantitative predictions are difficult~\cite{Kharzeev:2007jp,Toneev:2010ph,Chen:2019qoe,Shi:2017cpu}. 
To date, no quantitative conclusion on the CME has been reached at RHIC; an upper limit, as we report here, should provide significant insights into the CME at RHIC.

In order to isolate the resonance background contributions, we report measurements of the $\dg$ variable, 
differential in pair invariant mass ($\minv$). 
The integral $\Delta\gamma$ with a minimum $\minv$ limit is presented.
To fully exploit the data, an ESE~\cite{Schukraft:2012ah} technique is deployed 
to determine the $v_{2}$ background shape as a function of $\minv$.   
The $\dg(\minv)$ data are then fitted to the $v_{2}$ background shape 
plus a $\minv$-independent constant term. 
The extracted constant term represents a $v_{2}$-independent component in the data, possibly a $\minv$-integrated CME signal.

\section{Experiment and data analysis}
The data reported here were taken by the STAR experiment 
at the center-of-mass energy per nucleon pair of \sNN = 200 GeV
in the year 2011, 2014 and 2016. 
A total of 2.5 billion minimum-bias (MB) triggered events  were used in the analysis.  
The STAR apparatus is described in Ref.~\cite{Ackermann:2002ad}. 

The main detectors used in this analysis are the time projection chamber (TPC)~\cite{Anderson:2003ur,Ackermann:1999kc} and the time-of-flight (TOF) detector~\cite{Llope:2012zz}. 
Track trajectories are reconstructed from hits detected in the TPC; 
at least 10 points out of a possible maximum of 45 points are required for a valid track. 
The primary interaction vertex is reconstructed from those tracks. 
Events with primary vertices within 30~cm (year 2011) or 6~cm (years 2014, 2016) longitudinally 
and within 2 cm in the transverse plane from the geometrical center of the TPC are used. 
The event centrality is determined from the multiplicity of those charged particle tracks which are within pseudorapidity $|\eta| < 0.5$, and have distance of closest approach (DCA) to the primary vertex of less than 3 cm. 

Tracks used for the analysis are required to have at least 20 points used in track fitting, and DCA less than 1~cm.
The fraction of fit points out of the maximum allowed by the TPC geometry 
is required to be greater than 0.52 to avoid track splitting.
Particle momenta are determined by the track trajectories in the STAR magnetic field.
A minimum transverse momentum ($\pT>0.2$ \GeVc) is required to ensure that each  
track traversing the TPC can reach the TOF detector. 
The charged particles can be identified by their ionization energy loss (\textit{dE/dx}) in the TPC gas 
and their time of flight from the TOF detector. 
Pions are identified up to $\pT$ = 0.8 \GeVc\ with \textit{dE/dx}, and extended to $\pT$ = 1.8 \GeVc\ with the TOF.

This analysis uses the three-particle correlator: 
\begin{equation}
\gamma=\mean{\cos(\phia+\phib-2\phic)}/\vc,
\label{eq:gamma}
\end{equation}
where $\alpha$ and $\beta$ represent the pion index, and the average $\mean{...}$ runs over all triplets and over all events.
The azimuthal angle of the third particle, $\phic$, serves as a measure of $\psi_2$.
The imprecision in determining the $\psi_2$ by a single particle is corrected by the resolution factor, 
equal to the particle's elliptic flow anisotropy $\vc$.
Charged TPC tracks with $\pT$ from 0.2 to 2 \GeVc\ are used for particle $c$.
Two methods are used: 
1) sub-event method, the main method used in this analysis, where the $\alpha$, $\beta$ particles are from one half of the TPC ($-1<\eta<-0.05$ or $0.05<\eta<1$) and the particle $c$ is from the other half ($0.05<\eta<1$ or $-1<\eta<-0.05$)~\cite{Adamczyk:2012ku}; 
2) full-event method, where the $\alpha$, $\beta$ and $c$ particle are 
all taken from the pseudo-rapidity range $|\eta| < 1$~\cite{Abelev:2009ad,Abelev:2009ac}. 
The $\eta$ gap of 0.1 between the positive and negative pseudorapidity subevents is to suppress short-range correlations from quantum interference and Coulomb interaction, as well as detector-related effects such as track splitting~\cite{STAR:2021mii}. 
In order to identify resonance decay contributions, the $\dg$ correlator is studied as a function of 
$\minv$ of the $\alpha$ and $\beta$ particle pairs. 
The analysis loops over $\alpha$ and $\beta$ particles, and the $c$ particle is handled by the cumulant method~\cite{Borghini:2002vp}.

The systematic uncertainties are estimated for each run 
by varying the required minimum number of hit points from 20 to 15, 
and the DCA from 1.0 cm to 2.0 and 0.8 cm. 
In the full-event method, the $\eta$ gap used to determine $\vc$ via two-particle correlations is varied from 1 to 0.5 and 1.4~\cite{Adamczyk:2012ku,STAR:2019xzd}.
In the sub-event method, the $\eta$ gap between the two sub-events is varied from 0.1 to 0.3~\cite{Adamczyk:2013gw}. 
For each variation, the statistical fluctuation effect arising from the change in the data sample is subtracted. 
For each source when multiple variations are assessed, the systematic uncertainty is taken as the root mean square.
The systematic uncertainties from the above sources are added in quadrature for each dataset of the three runs. 
The three datasets are then combined assuming their systematic uncertainties are fully correlated. 

The pion purity in this analysis is approximately 98\%; the systematic uncertainty from particle identification is found to be negligible. The effect of different charge combinations among the three particles in Eq.~(\ref{eq:gamma}) has been studied in Ref.~\cite{Abelev:2009ad} and found to cause negligible difference in the $\dg$ results. 

For the extracted possible CME signal from the ESE fit method, a fit result is obtained for each of the above variations,
and the systematic uncertainty is estimated in the same way as described above.
The estimated absolute systematic uncertainties on the extracted CME signal fraction in the 20-50$\%$ centrality are 2$\%$, 3$\%$, and 3$\%$ for DCA, number of hit points, and $\eta$ gap variations, respectively.
The $\minv$ range used in the ESE fit is varied from above 0.4 \GeVcsq\ to above 0.35 and 0.45 \GeVcsq, which yields negligible change in the results.
Table~\ref{systable} summarizes the systematic uncertainties on the extracted possible CME signal fraction.
\begin{table}[!htpb]
\centering
\begin{tabular}{lccr}
    \hline
	DCA\hspace{3mm} & \hspace{1mm}Number of points\hspace{1mm} & \hspace{1mm}Sub-event $\eta$ gap\hspace{1mm} & \hspace{3mm}Total \\ 
    \hline
    $\pm 2\%$  & $\pm 3\%$ & $\pm 3\%$ & $\pm 5\%$ \\ 
\hline
\end{tabular}
	\caption{The absolute systematic uncertainties on the extracted possible CME signal fraction (as a percentage) in the inclusive $\dg$ in 20-50$\%$ centrality Au+Au collisions at 200 GeV from the ESE fit method.}
\label{systable}
\end{table}

\section{Results and discussions}
\subsection{$\dg$ as function of $\minv$}

\begin{figure}[hbt]
	\begin{center}
		\includegraphics[width=0.4\textwidth]{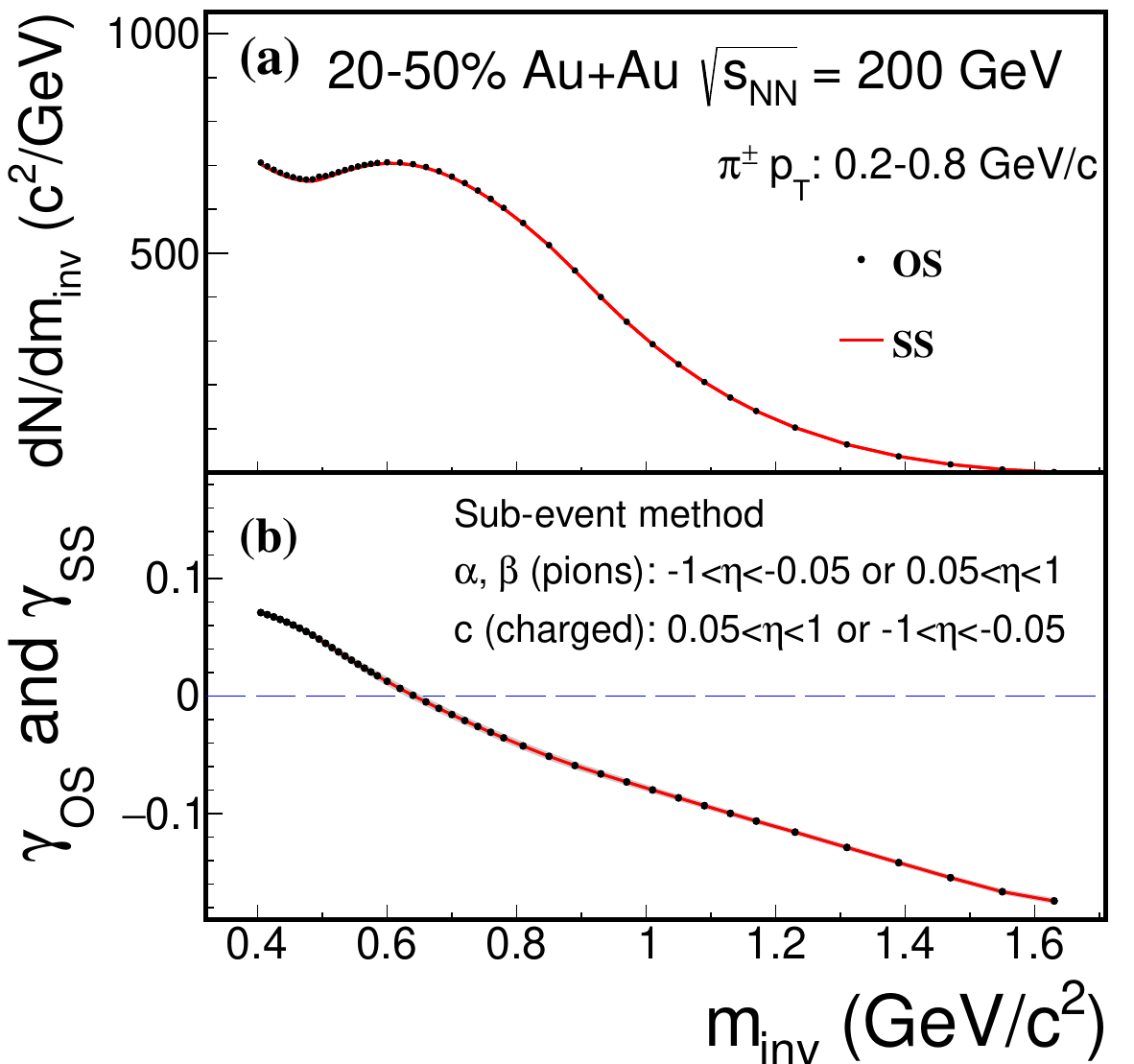}
		\caption{The $m_{\rm inv}$ dependences of (a) the OS and SS pion pair multiplicities, 
			and (b) $\gOS$, $\gSS$ in 20-50$\%$ Au+Au collisions at 200 GeV. 
			Error bars are statistical. The shaded areas in (b) are systematic uncertainties, which are small. 
			}
		\label{fig:OSSS}
	\end{center}
\end{figure}

Figure~\ref{fig:OSSS}(a) shows the number of OS and SS $\pi^+\pi^-$ pairs as functions of $\minv$ for 20-50$\%$ centrality Au+Au collisions at \sNN = 200 GeV. Pions are identified by the TPC $dE/dx$ method within $0.2<p_T<0.8$~\gevc. The sub-event method is used. The $N_{\rm OS}$ and $N_{\rm SS}$ are nearly identical. 
Figure~\ref{fig:OSSS}(b) shows the average correlators for OS and SS pairs, $\gamma_{\rm OS}$ and $\gamma_{\rm SS}$, respectively. The positive (negative) values at low (high) $\minv$ arise from pair kinematics coupled with particle $v_2$.
Low $\minv$ pairs tend to be close in azimuth and hence lead to positive $\gamma$ values, while high $\minv$ pairs tend to be back-to-back and hence lead to negative $\gamma$ values.
Because of these correlations between $\minv$ and the pair opening angle, the $\gamma_{\rm OS}$ and $\gamma_{\rm SS}$ vary over large range as a function of $\minv$ and look nearly identical. 
The integrated $\gamma_{\rm OS}$ and $\gamma_{\rm SS}$ (of pions in this analysis) over $\minv$ (effectively over opening angle) are close to zero and comparable with previous results of charged hadrons ~\cite{Abelev:2009ad,Abelev:2009ac}. 
These kinematics-related large variations in $\gamma_{\rm OS}$ and $\gamma_{\rm SS}$ over $\minv$ are removed in the difference of Eq.~(\ref{eq:dg}), which we present next.

\begin{figure}[hbt]
	\begin{center}
		\includegraphics[width=0.48\textwidth]{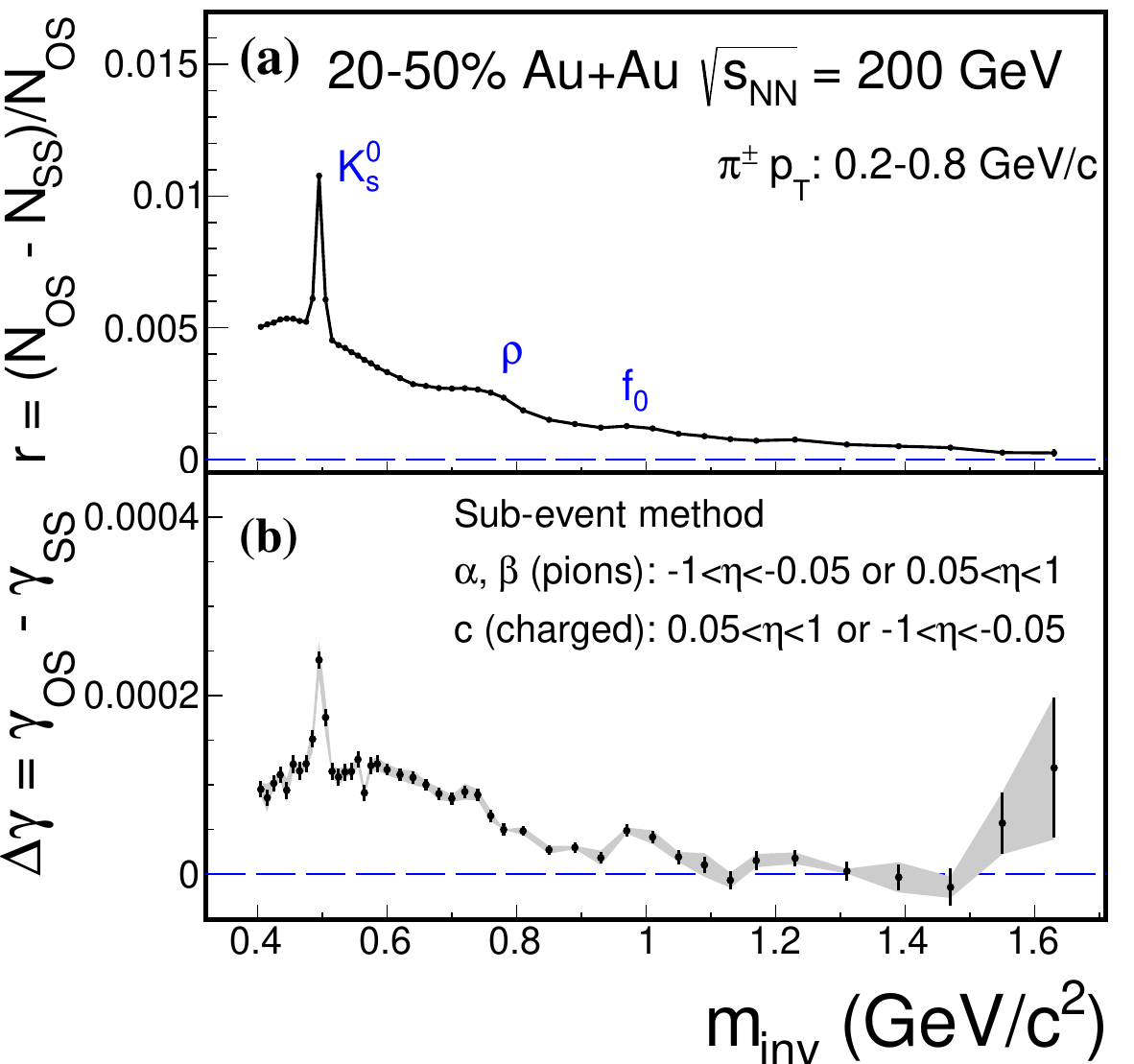}
		\caption{
			$m_{\rm inv}$ dependences of (a) the relative excess of OS over SS pion pairs, 
			and (b) $\dg=\gOS-\gSS $ in 20-50$\%$ Au+Au collisions at 200 GeV. 
			Error bars are statistical. The shaded areas in (b) are systematic uncertainties.
			}
		\label{fig:mass}
	\end{center}
\end{figure}

Figure~\ref{fig:mass}(a) shows the relative OS and SS $\pi^+\pi^-$ pair abundance difference, $r = (N_{\rm OS} - N_{\rm SS})/N_{\rm OS}$, 
as a function of $\minv$ for 20-50$\%$ centrality Au+Au collisions at \sNN = 200 GeV.
Figure~\ref{fig:mass}(b) shows the measured $\dg$ as a function of $\minv$ in a similar way. 
The $\minv<0.4$~\GeVcsq\ region is excluded because 
the acceptance difference between OS and SS pairs, mostly close in azimuthal angle, 
becomes unreasonably large.
This is due to the charge-dependent non-uniformity of the TPC acceptance/efficiency in the azimuthal direction; the correction become too large at small $\minv$. 
As shown in Fig.~\ref{fig:mass}(b), a clear peak from $K_{s}^{0} \rightarrow \pi^{+}+\pi^{-}$ decay is observed in $\dg$, 
and possible $\rho^{0}$ and $f_{0}$ peaks are also visible~\cite{Adams:2003cc}.
These peaks correspond to the resonance production peaks in $r$ shown in Fig.~\ref{fig:mass}(a). The results indicate strong contributions from resonances to the $\Delta\gamma$ observable.

It is clear from the comparison of the two panels in Fig.~\ref{fig:mass} that the inclusive $\dg$ measurement is contaminated by a large background caused by resonance decays and correlated particle pairs. The possible CME signal is in principle hidden within the large background in the $\dg$ measurement of Fig.~\ref{fig:mass}(b). 
This CME signal would be the difference in $\gamma_{\rm OS}$ and $\gamma_{\rm SS}$ inherited from initial correlations, rather than correlations from the final state, such as resonance decays.

\subsection{$\dg$ at large $\minv$}

As indicated by Fig.~\ref{fig:mass}(a), most of the excess of OS over SS pion pairs are from the small $\minv$ region (approximately 96$\%$ at $\minv<1$~\GeVcsq). 
Applying a minimum $\minv$ requirement would reduce those contributions. 
On the other hand, a lower $\minv$ cut would remove a large fraction of low $p_T$ pions.
It may thus reduce the possible CME signal because the CME is theoretically conjectured to be a low $\pT$ phenomenon~\cite{Kharzeev:2007jp,Abelev:2009ad}.
A recent study~\cite{Shi:2017cpu}, however, suggests a rather $\pT$ independent signal above 0.2 \GeVc.

\begin{figure}[hbt]
	\begin{center}
		\includegraphics[width=0.45\textwidth]{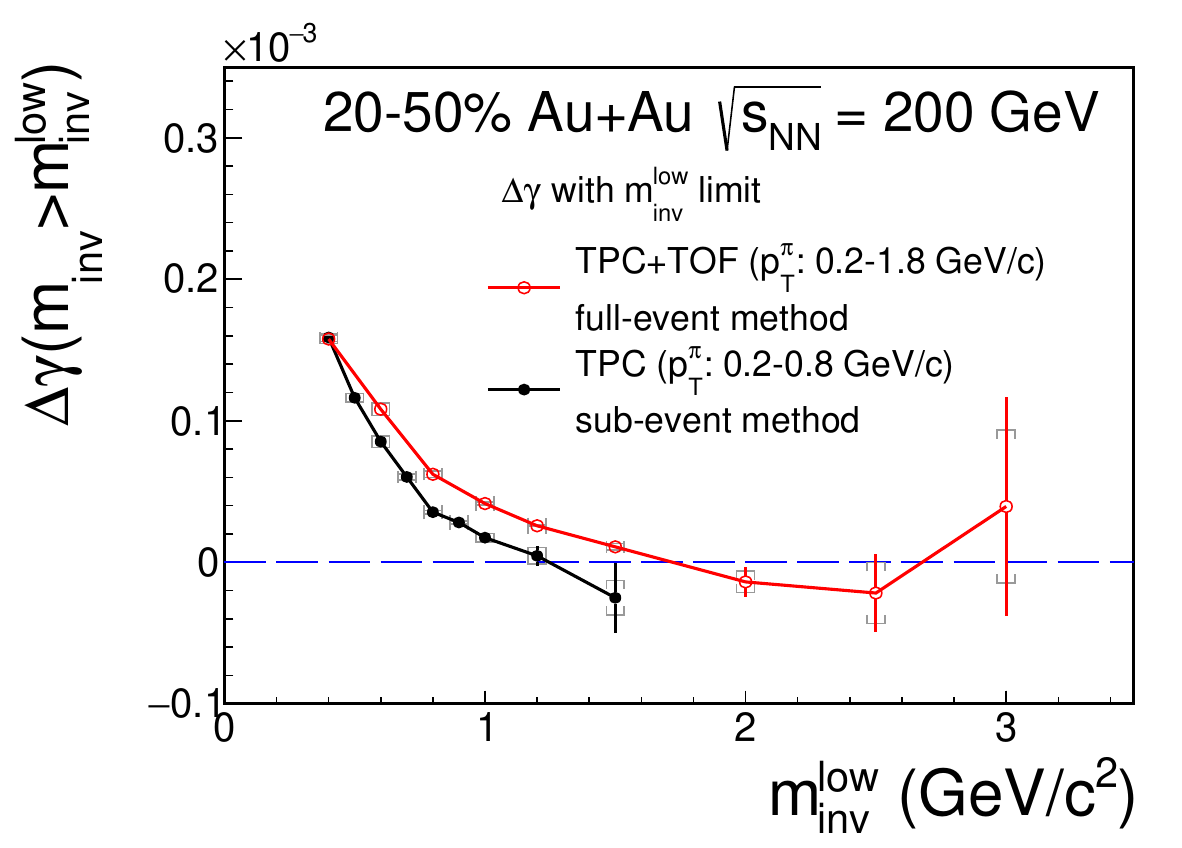}
		\caption{The $\pi$ pair $\dg$ at $m_{\rm inv} > m_{\rm inv}^{\rm low}$ 
for the data shown in Fig.~\ref{fig:mass} identified by TPC (black points) and for the more extended $p_T$ range identified by TPC and TOF (red circles).
			Error bars are statistical. 
			The caps are systematic uncertainties.
			}
		\label{fig:HM}
	\end{center}
\end{figure}

In any case, it is interesting to examine the $\Delta\gamma(\minv>m_{\rm inv}^{\rm low})$ above a certain $m_{\rm inv}^{\rm low}$ value, 
which would be more sensitive to the CME signal if the signal is a more slowly decreasing function of $\minv$ than resonance contributions. 
This is presented in Fig.~\ref{fig:HM}, where two results are shown. Black points show the $\Delta\gamma(\minv>m_{\rm inv}^{\rm low})$ for data in Fig.~\ref{fig:mass}. Because the pions are identified by the TPC up to  $p_T=0.8$~\gevc, the data do not reach a large enough $\minv$. 
In order to measure the spectra at large $\minv$ with high statistics, 
we include the pions identified by TPC and TOF detectors with $p_{T}$ extended to 1.8~\gevc, and the full-event method is used.
The result is shown in red circles.
The result from the full-event method with the higher $\pT$ range is systematically larger than that from the sub-event method with the limited $\pT$ range.
The reasons are two folds: the $\dg$ has been found to increase with $\pT$ \cite{Abelev:2009ad} and the $\eta$ gap in the sub-event method suppresses correlation background. For both cases, the $\Delta\gamma(\minv>m_{\rm inv}^{\rm low})$ decreases with increasing $m_{\rm inv}^{\rm low}$ and approaches zero when $m_{\rm inv}^{\rm low}$ becomes large. 
Note that residual resonance and other correlation backgrounds may still remain at high mass, 
and detailed model and theoretical studies are required to draw further conclusions.

\subsection{Event-shape engineering to extract CME}

In order to fully exploit the data to extract a possible CME signal over the entire $\minv$ range, resonance contributions need to be removed.
This may be achieved by taking advantage of the presumably different $\minv$ dependencies of the background and the possible CME signal. 
Assuming the $\Delta\gamma$ data contain the flow-induced background and a possible CME signal, 
the inclusive $\dg$ can be expressed as~\cite{Zhao:2017nfq} 
\begin{equation}
	\dg(\it m_{\rm inv}) = r(\it m_{\rm inv}) \rm \mean{\cos(\phi_{\alpha} + \phi_{\beta} -2\phi_{\rm res})} \it v_{\rm 2, res} + \rm \dg_{\rm sig}.
	\label{eq3}
\end{equation}
The first term of Eq.~(\ref{eq3}) r.h.s.~is the background and is obviously dependent on $\minv$ and $v_2$. 
The second term is the possible CME signal. 

\begin{figure}[hbt]
	\begin{center}
		\includegraphics[width=0.48\textwidth]{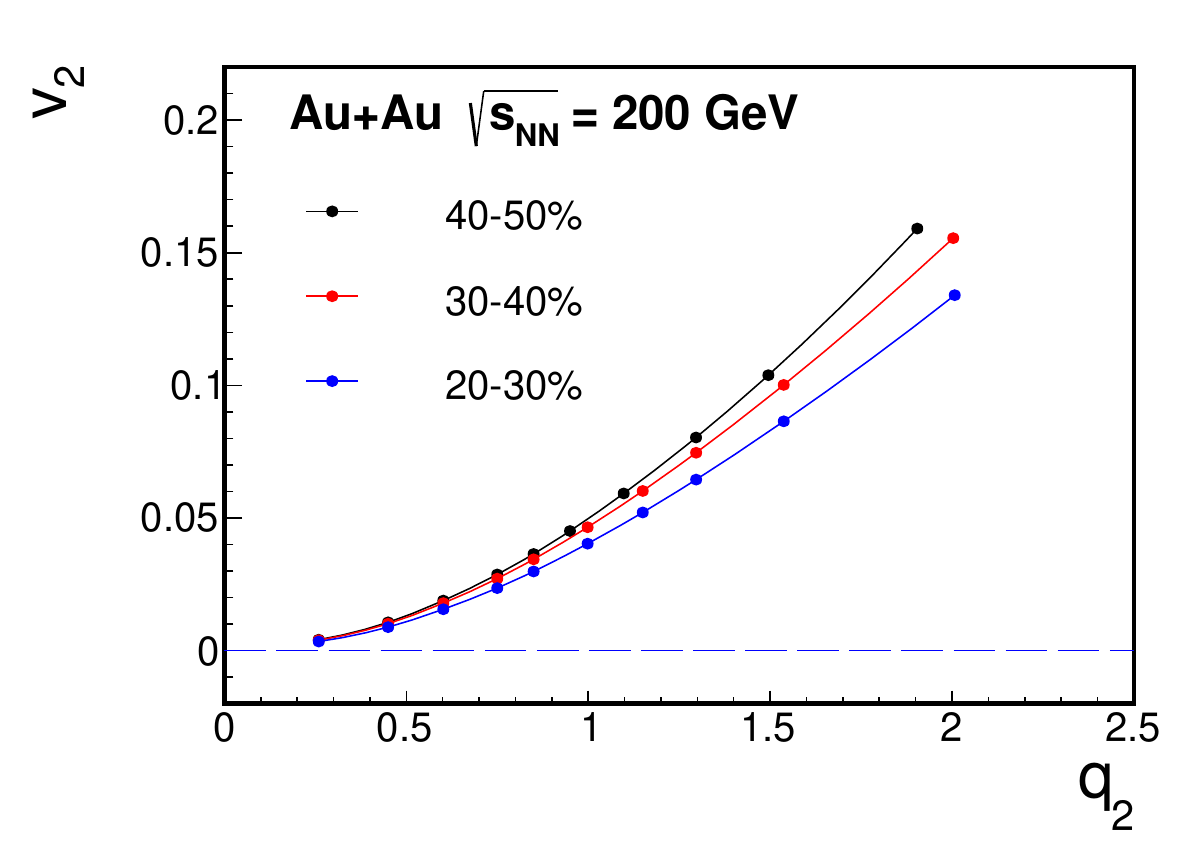}
		\caption{The elliptic flow $v_2$ in bins of $q_2$ for the three centrality bins in the 20-50$\%$ range of Au+Au collisions at 200 GeV.
		}
		\label{fig:q2v2}
	\end{center}
\end{figure}

We attempt to separate the background and the CME by resorting to the ESE method. The ESE method selects events from a narrow centrality bin
with different $v_{2}$ values by using the reduced flow vector $q_{2}$ quantity;
$q_{2} = |\sum_{j=1}^{N}{e^{i2\phi_{j}}}|/\sqrt{N}$ summing over the $\alpha, \beta$ particles in each event using the subevent method. 
The $v_2$ is calculated as a function of $q_2$ by the correlations w.r.t.~the particles from the other subevent.
Figure ~\ref{fig:q2v2} shows the $v_2$ in bins of $q_2$ in each narrow centrality bin.
The $v_{2}$ is strongly correlated with $q_{2}$ because
the $\alpha$, $\beta$ particles are used for both the $q_{2}$ and $v_{2}$ calculation.
In other words, this ESE method is selecting mainly on the statistical fluctuations of the $\alpha$/$\beta$ particle's elliptic anisotropy, which has a wide spread. The $v_2$ does not linearly depend on $q_2$, especially at small values of $q_2$. This has also been observed in simulations~\cite{Wen:2016zic}.

\begin{figure}[hbt]
	\centering
	\includegraphics[width=0.48\textwidth]{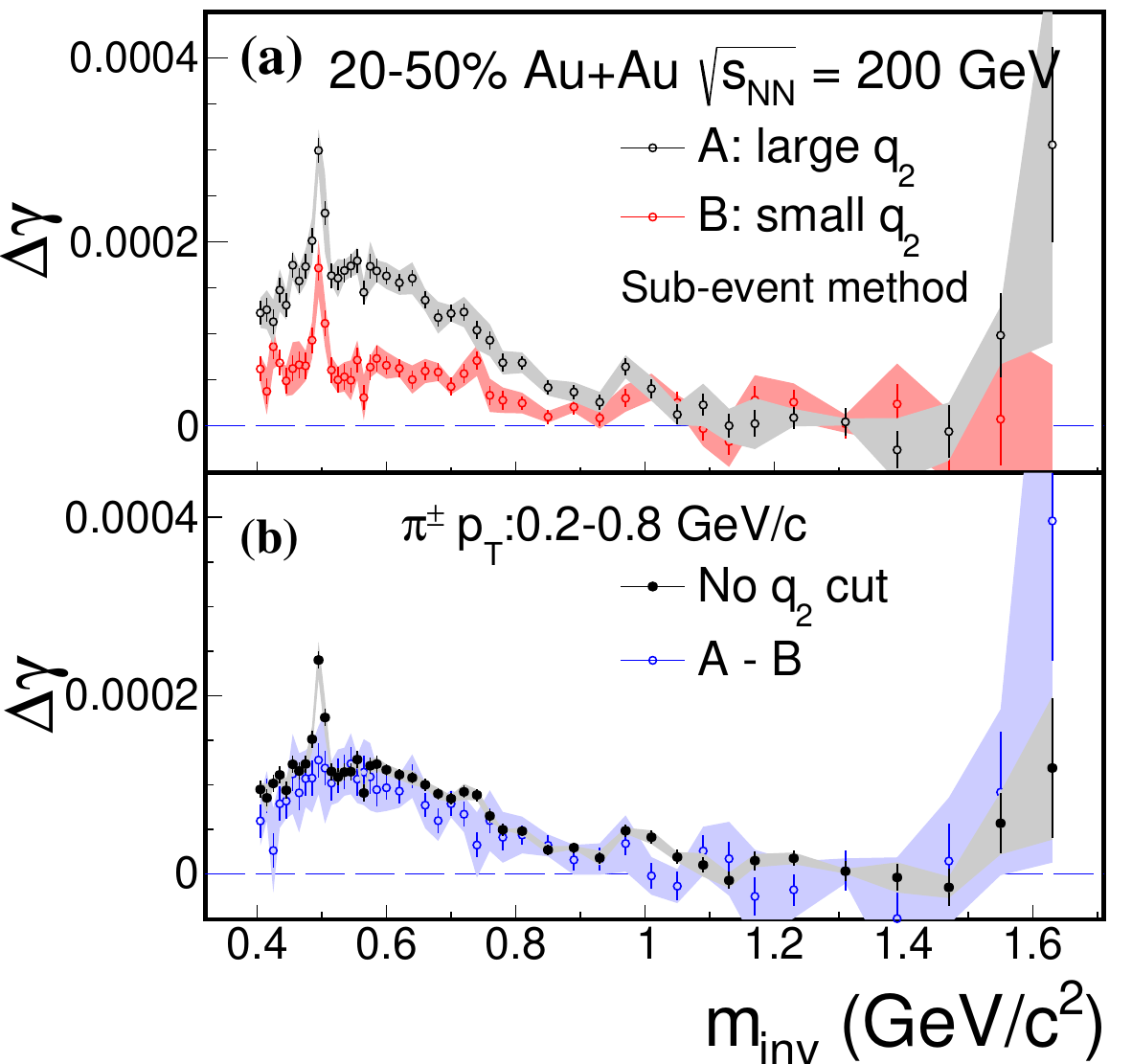}
	\caption{
		$m_{\rm inv}$ dependences of (a) the $\Delta\gamma$ from ESE-selected event samples A 
		(50$\%$ largest $q_{2}$) and B (50$\%$ smallest $q_{2}$), respectively,
		and (b) the inclusive (no $q_{2}$ restriction) $\dg$ compared with $\rm \dg_{A} - \dg_{B}$ in 20-50$\%$ Au+Au collisions at 200 GeV.
		Error bars are statistical. The shaded areas are systematic uncertainties.}
	\label{fig:q2}
\end{figure}

The events shown in Fig.~\ref{fig:mass} are divided into two equal-size groups according to the $q_{2}$ value in each narrow centrality bin: 
event sample A with the 50$\%$ largest $q_{2}$ and event sample B with the 50$\%$ smallest $q_{2}$, as indicated in Fig.~\ref{fig:q2v2}(a). 
The $q_2$ cut values to separate events are similar among the three centrality bins. The $\dg$ values are calculated in the two halves of events for each centrality bin, and then are combined.
Figure~\ref{fig:q2}(a) shows the $m_{\rm inv}$ dependence of the $\Delta\gamma_{\rm A}$ and $\Delta\gamma_{\rm B}$ from event samples A and B, 
respectively, integrated over the 20-50$\%$ centrality range.
Figure~\ref{fig:q2}(b) shows the inclusive (no $q_{2}$ restriction, i.e.~the same data as shown in Fig.~\ref{fig:mass}(b)) $\dg$ compared with $\dg_{\rm A} - \dg_{\rm B}$. 
The systematic uncertainty of the latter is larger than twice that of the former, which is approximately $(\Delta\gamma_{\rm A}+\Delta\gamma_{\rm B})/2$. This is due to an anticorrelation between the two event classes as they are selected largely on statistical fluctuations as aforementioned.

In order to exploit the presumably different $\minv$ dependencies of the background and the signal to extract the CME, we need to assess the background shape in $\minv$. 
To this end, we first have verified that the $r(\minv)$ distributions are the same between the two event classes. The decay angular correlations, $\mean{\cos(\phi_{\alpha} + \phi_{\beta} -2\phi_{\rm res})}$, are presumably also the same. 
It is probably safe to assume that the $v_2(\minv)$ has the same $\minv$ dependence for the different $q_2 $ event classes, only differing in magnitude. 
Under this assumption, the backgrounds in different $q_{2}$ event classes will have the same shape in $\minv$, differing only in its overall magnitude. 
The CME depends on the overall magnetic field. As mentioned in the introduction, the magnetic field is primarily produced by spectator protons. Since the spectator protons and the participant particle anisotropy are uncorrelated~\cite{Xu:2017qfs}, the CME is largely independent of the $v_2$ of the different $q_2$ classes. However, the participant protons do contribute to the overall magnetic field~\cite{Bzdak:2011yy,Belmont:2016oqp,Sun:2019hao}. Model calculations indicate that such a contribution could account for 20\% of the overall magnetic field strength 
in Au+Au collisions of medium centrality~\cite{Sun:2019hao}.
This magnetic field contribution could have variations over the different $q_2$ event classes due to the presumably different event shape selected by $q_2$. 
In this paper, we assume such variation is small and proceed with the assumption that the CME is 
independent of $q_2$ and $v_2$
Then the difference of the $\dg(m_{\rm inv})$ from the different $q_{2}$ event classes can be regarded as  
the background $\dg_{\rm bkgd}(m_{\rm inv})$ shape~\cite{Zhao:2018ixy}.

\begin{figure}[hbt]
	\centering
	\includegraphics[width=0.48\textwidth]{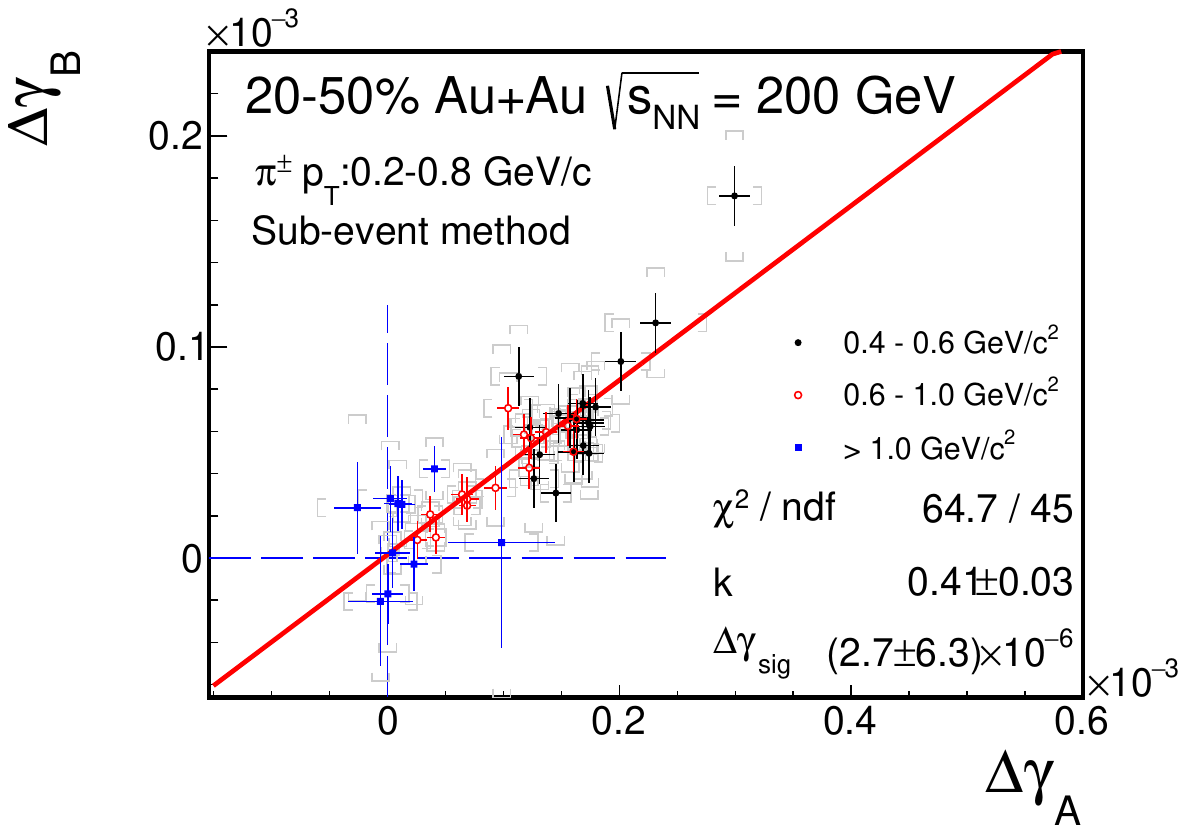}
	\caption{$\dg_{\rm A}$ versus $\dg_{\rm B}$ in 20-50$\%$ Au+Au collisions at 200 GeV, with the linear function fit of Eq.~(\ref{eq4}). 
	Error bars are statistical. Horizontal and vertical caps are the systematic uncertainties on $\dg_{\rm A}$ and $\dg_{\rm B}$.
	Marker colors indicate the data from different $\minv$ regions (black: 0.4-0.6 GeV/$c^2$, red: 0.6-1.0 GeV/$c^2$, blue: $>1.0$ GeV/$c^2$).
	} 
	\label{fig:fit}
\end{figure}

The inclusive $\dg$ contains both the background and the possible CME. 
We further assume that the possible CME signal is independent of $\minv$.
Then, with the background shape given by $\dg_{\rm A}-\dg_{\rm B}$, 
the possible CME signal can be extracted from a two parameter fit: $\dg=b(\dg_{\rm A}-\dg_{\rm B})+{\rm \dg_{\rm sig}}$. 
However, since the same data are used in $\dg$ and $\dg_{\rm A}-\dg_{\rm B}$, their statistical errors are not independent. 
To properly handle statistical errors, an alternative function is used to fit the two independent measurements of $\dg_{\rm A}$ versus $\dg_{\rm B}$,
namely: 
\begin{equation}
	\dg_{\rm B}=k\dg_{\rm A}+(1-k){\rm \dg_{\rm sig}}, 
	\label{eq4}
\end{equation}
where $k$ and ${\rm \dg_{\rm sig}}$ are the fit parameters.
Since $\dg\approx(\dg_{\rm A}+\dg_{\rm B})/2$, then $b\approx(1+k)/(1-k)/2$.
In this fit model, the background is not required to be strictly proportional to $v_2$, but only dependent on $v_2$~\cite{Zhao:2017nfq,Li:2018oot}. 

Figure~\ref{fig:fit} shows $\dg_{\rm A}$ versus $\dg_{\rm B}$ in 20-50$\%$ centrality Au+Au collisions at 200 GeV. 
Each data point corresponds to one $\minv$ bin in Fig.~\ref{fig:fit}(a).
The line is the fit by Eq.~(\ref{eq4}).
The fitted ${\rm \dg_{\rm sig}}$ is (0.03 $\pm$ 0.06 $\pm$ 0.08) $\times10^{-4}$ and is found to be $(2\pm4\pm5)$~\% of the inclusive $\Delta\gamma(\minv>0.4$ GeV/$c^2)=(1.58 \pm 0.02 \pm 0.02) \times10^{-4}$. 
These values represent over an order of magnitude reduction 
from the inclusive $\dg$ measurement. 
Our results indicate that the possible CME signal is small in the inclusive $\Delta\gamma$, consistent with zero with current precision. 
This presents an upper limit of $0.23\times10^{-4}$, or $15\%$ of the inclusive result at $95\%$ confidence level~\cite{Feldman:1997qc}. 

We note, as previously discussed, that our two-component fit model is based on the following assumptions: (1) the $v_2(\minv)$ dependence is the same between the two $q_2$ event classes; (2) the magnetic field contribution from participant protons has negligible variation between the two $q_2$ event classes; and (3) the CME signal $\dg_{\rm sig}$ is independent of $\minv$.
The $\chi^2/$ndf of our fit in Fig.~\ref{fig:fit} indicates a $p$-value of 2.86\%, which suggests that our assumptions 
may be reasonable, but future improvement with the help of theoretical calculations is possible.
Nevertheless, the fitted $\Delta\gamma_{\rm sig}$ may be interpreted as the signal averaged over the $\minv$ range.
The potential CME could depend on $\minv$, 
and given enough statistics, such details could be investigated experimentally by more sophisticated ESE analysis.

Our result is consistent with our previous finding from an ESE analysis~\cite{Adamczyk:2013kcb} 
and a more recent measurement using spectator and participant planes~\cite{STAR:2021pwb}. 
The recent isobar data~\cite{STAR:2021mii} do not yield an observable CME signal, in line with the present work.
Our upper limit is quantitatively similar to those reported at the LHC~\cite{Sirunyan:2017quh,Acharya:2017fau}.
Quantitative predictions of the CME signal strength are not available at RHIC or the LHC; only a general expectation of $10^{-4}$ has been suggested~\cite{Kharzeev:2004ey,Shi:2017cpu}. Our upper limit, together with those at the LHC, add significant insights to the physics of the CME, and calls for further theoretical inputs.

\section{Conclusions}
In summary, we report differential measurements of the reaction-plane-dependent azimuthal correlation of pion pairs ($\Delta\gamma$), sensitive to the topological-charge-induced chiral magnetic effect in QCD, as a function of the pair invariant mass ($\minv$).
Resonance structures are observed in $\dg$($\minv$), indicating the dominance of background contributions in the previous inclusive $\dg$ measurements~\cite{Abelev:2009ad,Abelev:2009ac,Adamczyk:2014mzf}. 
At large $\minv$, where this background is significantly reduced, the $\Delta\gamma$ is also significantly smaller.
To isolate the possible CME signal from background, 
event shape engineering by the sub-event method is used to determine the background shape in $m_{\rm inv}$.
The background shape is used in a 
two-component fit to the $\dg$($m_{\rm inv}$) data, 
assuming it contains a $v_{2}$-independent signal in additional to the $v_{2}$-dependent background. 
Such a fit yields a $v_{2}$-independent signal of $\dg_{\rm sig}$ = (0.03 $\pm$ 0.06 $\pm$ 0.08) $\times10^{-4}$ in 20-50$\%$ centrality Au+Au collisions at 200 GeV,
$(2\pm4\pm5)\%$ of the inclusive measurement of $\dg$($\minv>0.4$ GeV/$c^2$)=(1.58 $\pm$ 0.02 $\pm$ 0.02) $\times10^{-4}$, 
within pion $\pT=0.2-0.8$ \GeVc\ and averaged between pseudorapidity ranges of $-1<\eta<-0.05$ and $0.05<\eta<1$. 
This represents an upper limit of $0.23\times10^{-4}$, or $15\%$ of the inclusive result,
at $95\%$ confidence level for the possible CME signal integrated over $\minv$. 
This constitutes a report of an upper limit on the theoretically predicted CME at RHIC, with explicit isolation of background and under several assumptions. Theoretical inputs on the $m_{\rm inv}$ dependence of the CME as well as magnetic field calculations will be helpful to improve this upper limit in the future.

\section*{Acknowledgments}
We thank the RHIC Operations Group and RCF at BNL, the NERSC Center at LBNL, and the Open Science Grid consortium for providing resources and support.  This work was supported in part by the Office of Nuclear Physics within the U.S. DOE Office of Science, the U.S. National Science Foundation, National Natural Science Foundation of China, Chinese Academy of Science, the Ministry of Science and Technology of China and the Chinese Ministry of Education, the Higher Education Sprout Project by Ministry of Education at NCKU, the National Research Foundation of Korea, Czech Science Foundation and Ministry of Education, Youth and Sports of the Czech Republic, Hungarian National Research, Development and Innovation Office, New National Excellency Programme of the Hungarian Ministry of Human Capacities, Department of Atomic Energy and Department of Science and Technology of the Government of India, the National Science Centre of Poland, the Ministry of Science, Education and Sports of the Republic of Croatia, German Bundesministerium f\"ur Bildung, Wissenschaft, Forschung and Technologie (BMBF), Helmholtz Association, Ministry of Education, Culture, Sports, Science, and Technology (MEXT) and Japan Society for the Promotion of Science (JSPS).
%

\bibliographystyle{unsrt}
\bibliography{ref}

\end{document}